\documentclass[aps,prc,preprintnumbers,epsf,nofootinbib]{revtex4}
\usepackage[dvips]{graphicx}
\newcommand{\bea}{\begin{eqnarray}}
\newcommand{\eea}{\end{eqnarray}}

\begin{document}
\preprint{arXiv: 0711.4637v6[nucl-th]}
\title{A nonperturbative parametrization and scenario for EFT renormalization}
\author{Ji-Feng Yang}
\address{Department of Physics, East China Normal University,
Shanghai, 200062, China}
\date{March 10, 2009}
\begin{abstract}
We present a universal form of the $T$-matrices renormalized in
nonperturbative regime and the ensuing notions and properties that
fail conventional wisdoms. A universal scale is identified and shown
to be renormalization group invariant. The effective range
parameters are derived in a nonperturbative scenario with some new
predictions within the realm of contact potentials. Some
controversies are shown to be due to the failure of conventional
wisdoms.
\end{abstract}
\maketitle

Applications of the effective field theory (EFT) methods now prevail
in physical literature. In particular, the applications of the EFT
approach to nucleon systems has been producing many encouraging
results\cite{BvKERev}, pointing towards a promising field
theoretical framework for nuclear system. However, the
nonperturbative nature makes the renormalization of such EFT's
rather nontrivial and still creates
controvsersies\cite{NTvK,PVRAB,EpelMeis} to be settled. Sufficient
evidences have been accumulated that the conventional wisdoms for
renormalization cease to apply straightforwardly in nonperturbative
regimes. This is not unexpected as they are established within
perturbative frameworks. Therefore it is desirable to reveal novel
notions and aspects of renormalization that deviate from the
conventional wisdoms. In particular, it is desirable to obtain more
concrete parametrization of the prescription dependence of the
objects (here, the $T$-matrix) in nonperturbative regimes as much as
possible.

To this end, we work out the rigorous solutions of the $T$-matrices
for low-energy nucleon-nucleon ($NN$) scattering that solve the
Lippmann-Schwinger equation (LSE) in all partial wave channels,
using contact potentials constructed according to the chiral EFT
approach\cite{WeinEFT}. The $^1S_0$ channel has been worked out in
Ref.\cite{C71} up to chiral order $\Delta=4$. Here, we present the
universal forms for both uncoupled and coupled channels. Then it is
immediate to see some features and notions that are intrinsically
nonperturbative and deviate from the conventional wisdoms of
renormalization within perturbative frameworks. These are important
conceptual gainings that could resolve most of the controversies
about the applications of EFT in nonperturbative regimes. The
notions and scenario demonstrated below are naturally illuminating
for any problem that is beset with nonperturbative divergences,
especially in the systems governed by singular short-distance
interactions. In addition, the analytical results and scenario
presented here could also be seen as a field-theoretical solution to
the universality of large scattering lengths in atomic and molecular
systems\cite{Braaten-Hammer}.

According to Weinberg\cite{WeinEFT}, the EFT approach to $NN$
scattering consists of two steps: First, the potentials for $NN$
scattering are systematically constructed using chiral perturbation
theory ($\chi$PT) as the relevant EFT up to certain chiral order
$\Delta$: ${\mathcal{O}}(\{p,m_\pi\}^{\Delta}/\Lambda^{\Delta})$
with $\Lambda$ being the upper limit for low energy $NN$ scattering
(e.g., $\Lambda\sim0.5\texttt{GeV}$); Second, the nonperturbative
$NN$ scattering $T$-matrices could be obtained by solving the LSE
with the potential constructed in the previous step. As LSE is hard
to solve rigorously for pion-exchange potentials, we first work with
contact potentials (EFT($\not\!\pi$)) that facilitate rigorous
solutions where the LSE for $L$-wave reduces to an algebraic form by
employing the following factorization trick or
ansatz\cite{Phillips}: \bea &&V\equiv
q^L(q^\prime)^L\sum_{i,j=0}^{\Delta/2-L}\lambda_{ij}
q^{2i}(q^\prime)^{2j}=q^L(q^\prime)^LU^T(q)\lambda
U(q^\prime),\nonumber\\
&&T\equiv q^L(q^\prime)^L\sum_{i,j=0}^{\Delta/2-L}\tau_{ij}
q^{2i}(q^\prime)^{2j} =q^L(q^\prime)^LU^T(q)\tau U(q^\prime),\eea
with $U(q)\equiv(1,q^2,q^4,\cdots,q^{\Delta-2L})^T$ being a column
vector and $q,q^\prime$ being the external off-shell momenta. This
is because for pionless interactions, the $NN$ potential up to order
$\Delta$ degrades into contact interactions that become a polynomial
in terms of external momenta up to power $\Delta$. Here $\lambda$
denotes the real symmetric matrix comprises of the contact couplings
($[C_{\cdots}]$) as $V$ is symmetric with respect to the external
momenta $q$ and $q^\prime$. For example, in $^1S_0$ channels at
$\Delta=2$, we have, $$\lambda=\left(\begin{array}{cc}C_0 & C_2\\
C_2 & 0 \\\end{array}\right).$$

Then the algebraic LSE takes the following form:\bea
\tau(E)=\lambda+\lambda{\mathcal{I}}(E)\tau(E),\eea where the matrix
${\mathcal{I}}(E)$ comprises of the integrals arising from
convolution. Note that $\tau$ and ${\mathcal{I}}$ are symmetric
matrices as $\lambda$ is. As ${\mathcal{I}}$ is complex, so is
$\tau$. A general element of ${\mathcal{I}}$ can be parametrized as
follows\bea{\mathcal{I}}_{i,j}(E)&&\equiv \left\{\int
\frac{d^3k}{(2\pi)^3}\frac{k^{2(L+i+j-2)}}{E-k^2/M+i\epsilon}\right\}_R=
\sum_{m=1}^{L+i+j-2}
J_{2m+1}p^{2(L+i+j-m-2)}-{\mathcal{I}}_0p^{2(L+i+j-2)},\\
\label{I0}\mathcal{I}_0&&\equiv J_0+i\frac{M}{4\pi}p,\eea where the
subscript $R$ denotes any possible regularization and/or
renormalization prescription rendering the integrals finite and
$[J_0,J_{2m+1}]$ being the corresponding parametrization. This
algebraic LSE is easy to solve. Then, for any uncoupled partial wave
channel, the on-shell $T$-matrices could be readily obtained from
$T=p^{2L}U^T(p)\tau U(p)$ with $p=\sqrt{ME}$ \cite{C71,3s1-3d1},
which can be simplified to the following form:\bea \label{InvT-ll}
\frac{1}{T_L}=\mathcal{I}_0 +\frac{
N_L([C_{\cdots}],[J_{2m+1}],p^2)}{
D_L([C_{\cdots}],[J_{2m+1}],p^2)p^{2L}},\eea where $N_L$ and $D_L$
are polynomials in terms of the following real parameters: the
couplings $[C_{\cdots}]$, the constants $[J_{2m+1},m>0]$ and $p^2$.
While for coupled channels, the inverse of the on-shell $T$-matrices
would take the following form:\bea \label{TINV}{\bf
T}^{-1}_J={\mathcal{I}_0}{\bf I} + {\bf \Delta}_J,\ \ \ \ {\bf
I}\equiv \left(
\begin{array}{cc} 1& 0   \\ 0 &1\\  \end{array}
\right),\ \ {\bf \Delta}_J\equiv \left(
\begin{array}{cc}
\frac{N_{J-1,J-1}}{D_{J-1,J-1}p^{2(J-1)}} ,& \frac{-
N_{J-1,J+1}}{D_{J-1,J+1}p^{2J}}
\\ \frac{-N_{J-1,J+1}}{D_{J-1,J+1}p^{2J}}, & \frac{
N_{J+1,J+1}}{D_{J+1,J+1}p^{2(J+1)}}\\
\end{array}
\right).\eea Again $[N_{\cdots},D_{\cdots}]$ are real polynomials in
terms of $[C_{\dots}]$, $[J_{2m+1},m>0]$ and $p$, hence (chiral)
perturbative in nature. Note that unitarity is automatically
satisfied here. At this stage, both the real part of
${\mathcal{I}_0}$ and the constants $[J_{2m+1},m>0]$ are
prescription dependent. The overall factors of $p^{2\cdots}$ have
been factored out so that the expansion of $[N_{\cdots},D_{\cdots}]$
in terms of $p^2$ starts from $p^0$. For example, the results for
$^3S_1-$$^3D_1$ at chiral order $\Delta=4$ read, \bea
\label{InvT-SD}&& {\bf T}^{-1}_{^3S_1- ^3D_1}={\mathcal{I}_0}{\bf
I}+{\bf \Delta}_{^3S_1- ^3D_1}, \ \ \ \ {\bf \Delta}_{^3S_1-
^3D_1}\equiv \frac{1} {{\mathcal{D}}_1p^4} \left(
\begin{array}{cc}
{\mathcal{N}}_1p^4, & - {\mathcal{D}}_{sd}p^2
\\- {\mathcal{D}}_{sd}p^2, & {\mathcal{D}}_0\\
\end{array}
\right)\eea where $ D_{0,2}=D_{2,2}=D_{0,0}$ and for convenience we
introduced the following notations: ${\mathcal{N}}_1\equiv N_{0,0},\
\ {\mathcal{D}}_1\equiv D_{0,0},\ \ {\mathcal{D}}_{sd}\equiv
N_{0,2},\ \ {\mathcal{D}}_0\equiv N_{2,2}$. For details we refer to
a forthcoming report\cite{3s1-3d1}.

Eqs.(\ref{InvT-ll}) and (\ref{TINV}) exhibit the following important
features: (1) First, the same complex parameter $\mathcal{I}_0$
appears in all channels in the same isolated position in $1/T$ or
${\bf T}^{-1}$ and the rest parts of $1/T$ or ${\bf T}^{-1}$ are
independent of $\mathcal{I}_0$, i.e., $\mathcal{I}_0$ is "decoupled"
from $[C_{\cdots}]$ and $[J_{2m+1},m>0]$ in every
channel\footnote{The rigorous proof of this point for $^1S_0$
channel has been given in Ref.\cite{C71}, which could be generalized
to higher channels, we will give the detailed proof in a forthcoming
report.}. This structure is most pivotal. (2) Second, with the
potentials truncated at finite order, only finite many of
$[J_{\cdots}]$ (or finite types of divergences) enter the game, in
spite that there are formally infinite many divergent items in the
iteration of LSE. (3) Both $[N_{\cdots}]$ and $[D_{\cdots}]$ are
chiral perturbative (or perturbative in the corresponding EFT).

Since the $p(=\sqrt{ME})$-dependence of the on-shell $T$-matrices
(and hence the inverse of $T$-matrices) is physical, the
prescription variations (i.e., variations of $[J_{\cdots}]$) must be
compensated by that of the couplings. This is nothing else but the
general principle of renormalization group (RG) invariance, then
appropriate combinations of the coefficients of $p$ in
$[N_{\cdots}]$ and $[D_{\cdots}]$ must be RG invariants. The most
outstanding point is that, the isolation or "decoupling" of
$\mathcal{I}_0$ from $[C_{\cdots}]$ and $[J_{2m+1},m>0]$ makes
itself a renormalization group invariant parameter in all channels,
hence, $J_0$ is a physical scale\cite{C71}. Therefore,
$\mathcal{I}_0$ is in fact a fundamental and universal parameter in
the low energy $NN$ scattering, and $J_0$ is no longer an ordinary
renormalization scale. Such a RG invariant quantity was also
predicted in Wilsonian RG approach\cite{Birse}, $\hat{V}_0$, whose
inverse is just $-\text{Re}({\mathcal{I}_0})=-J_0$ computed in the
Wilsonian cutoff approach. This is not the only deviation from the
conventional wisdoms for renormalization, according to which a
divergent integral usually produces sliding scales that are
physically meaningless within perturbative formulation.

In perturbative framework, the couplings in the contact potential
would get renormalized and "run". At lowest order, similar notion is
feasible in $^1S_0$ channel\cite{KSW}: $
1/T=J_0+iMp/(4\pi)+1/C_0=iMp/(4\pi) +M\mu/(4\pi)+1/C_{0;R}(\mu)$,
where $J_0+1/C_0=M\mu/(4\pi)+1/C_{0;R}(\mu)$ is RG invariant as
$N_0$ is a constant here, actually $N_0=1$. But at higher orders
(e.g., $\Delta\geq4$), it is easy to see that the rational
dependence of $T$-matrices upon $p$ precludes the conventional
wisdoms from being feasible, i.e., it is no longer possible to let
the variations in the prescription parameters $[J_{\cdots}]$ be
readily absorbed into the couplings and let the couplings "run".
This point could be seen from the requirement that the appropriate
combinations of the coefficients of $p$ in $[N_{\cdots},D_{\cdots}]$
should be RG invariant, which in turn imposes strong constraints
upon the variations in the couplings $[C_{\cdots}]$ and $[J_{2m+1},
m>0]$. ($J_0$ is already excluded from the set of prescription
parameters as it is an RG invariant scale now.) In fact, for $^1S_0$
at $\Delta=4$, the coefficients of the highest power term in $N_0,
D_0$ are respectively $N_{0;2}=C_4^2J_3^2, D_{0,3}=-C^2_4J_3$, which
could not make the ratios $N_{0;2}/N_{0;0}$ and $D_{0;3}/N_{0;0}$ RG
invariant at the same time, see Ref.\cite{C71}.

The second feature noted above engenders a novel notion of
"finiteness": Only a {\em finite} number of nonperturbative
divergences are to be removed in a manner preserving the functional
dependence upon $[C_{\cdots}]$ and $p$, which underlies the
feasibility of renormalization with a few nonperturbative
counterterms in Refs.\cite{YPhillips,EPVRAM}. This 'finiteness' is a
measure of nonperturbative renormalizability and not directly linked
to EFT power counting. Hence nonperturbative counterterms (termed as
"endogenous" in Ref.\cite{C71}) are not proportionate to
perturbative ones that obey EFT power counting. This is in sheer
contrast with perturbative renormalization programs, where
consistency requires that the counterterms be introduced or
constructed at exactly the same perturbation order as the divergent
vertex in consideration. This is another place where the
conventional wisdoms fail, usually interpreted as the
"inconsistency" of Weinberg's power counting. We will return to this
point later. This "finiteness" also underlies the feasibility of the
finite cutoff approaches\cite{Lepage,Rho,EGM,EMach,Bogner}.

In fact the nonperturbative form of $T$-matrices and their RG
invariance lead to an entanglement between the couplings and the
prescription: They must be defined coherently in order to match
physical boundaries. Then, the prescription must be appropriately
defined after the couplings are given first. Below, to obtain
unnaturally large scattering lengths and naturally sized effective
range, etc., we suggest a simple and natural strategy: The original
EFT($\chi$PT) power counting for potential construction are kept
intact, i.e., no modification of the power counting rules of the
couplings $[C_\Delta]$; In the meantime, $J_0$ and $[J_{2m+1},m>0]$
are so determined that physical boundary conditions are fulfilled.
Then, to yield large (unnatural) $S$-wave scattering lengths, the
most 'natural' or simplest scenario would be as follows\cite{C71}:
\bea \label{npt-PC} C_{\Delta}\sim 4\pi/(M\Lambda^{\Delta+1});\ \ \
J_0\sim M\Lambda/(4\pi) \sim |1/C_0|;\ \ \ J_{2m+1}\sim
M\mu^{2m+1}/(4\pi),\ m>0, \eea with $C_0$ being an $S$-wave contact
coupling at lowest order and $\mu$ of order $m_\pi$ or
$\sim100\texttt{MeV}$. In a generic EFT, $\mu\ll\Lambda$. Thus, the
only difference is with $J_0$. This is a "natural" scenario or
choice as $J_0$ is actually a fundamental and physical constant in
the nonperturbative regime, no longer an ordinary renormalization
scale. We will discuss other schemes in future works\cite{3s1-3d1}.

With the foregoing preparations, we could examine some important
theoretical issues in $NN$ low energy scattering. First, let us
consider some theoretical predictions. To this end, we calculate
effective range expansions (ERE) of $1/T$ in various channels. We
should remind that the following discussions are valid for contact
potentials only, not directly applicable to the cases containing
long range potentials such as the pion-exchange potentials for $NN$
systems. Later we will consider some speculations about such cases.

Let us start with the uncoupled cases, where the general form of ERE
of $1/T$ reads: \bea \label{ERE-general} \texttt{Re}\left
\{-\frac{4\pi}{M}\frac{
p^{2L}}{T_L}\right\}|_{p\rightarrow0}=p^{2L+1}\cot(\delta_L(p))
|_{p\rightarrow0} =-\frac{1}{a}+ \frac{1}{2}r_{e}p^2+
\sum_{k=2}^{\infty} v_{k}p^{2k},\eea with $a, r_{e}$ and $[v_{k},
k\geq2]$ being functions of the couplings in corresponding channels
and $[J_0,J_{2m+1},m>0]$. However, unlike the rest of
$[J_{\cdots}]$, $J_0$ contributes in each channel to only one of the
ERE parameters $\{a,r_{e},v_{k}, k\geq2\}$ that is the coefficient
of $p^{2L}$! This is obviously due to the special status of the
fundamental parameter ${\mathcal{I}_0}$. Employing the scenario
(\ref{npt-PC}) we could qualitatively deduce that, all but one ERE
parameters are naturally sized!  The exceptional one {\em might} be
unnaturally sized just because of the contribution of $J_0$. The
mechanism is simply that $$J_0+\frac{d^L
(\frac{N_L}{D_L})}{L!(dp^2)^L}|_{p=0}\sim
\frac{M}{4\pi}\mathcal{O}(\mu),$$ provided that the sign of
$\frac{d^L (N_L/D_L)}{L!(dp^2)^L}|_{p=0}$ is opposite to that of
$J_0$ as closer analysis shows that $\frac{d^L
(N_L/D_L)}{L!(dp^2)^L}|_{p=0}$ is of the same magnitude as $J_0$.
These general conclusions are summarized in Table I. It is known
that in $^1S_0$ channel the scattering length is unnaturally large
with the rest of ERE parameters being natural. Now, within the
context of contact potentials, higher ERE parameters might also be
unnaturally sized in an appropriate channel. These are new
predictions.

While in the coupled channels, one could find from Eq.(\ref{TINV})
that the diagonal entries of $\bf T$ take the following form:\bea
\frac{1}{T_L}=\mathcal{I}_0 +
\frac{{\mathcal{N}}_{L;0}+{\mathcal{I}_0}{\mathcal{N}}_{L;1}}
{{\mathcal{D}}_{L;0}+{\mathcal{I}_0}{\mathcal{D}}_{L;1}}p^{-2L},\eea
where $J_0$ enters into the rational terms, and hence precluding a
clear naturalness picture of the ERE parameters. However, using the
scenario of (\ref{npt-PC}) and the detailed contents of
$[{\mathcal{N}}_{\cdots},{\mathcal{D}}_{\cdots}]$, one could still
arrive at modestly good judgements, the status of naturalness in the
coupled channels is basically similar to that given in Table I.
There might be some deviations as $J_0$ now enters the rational
terms, which would affect the status of some ERE parameters. But
such influence would not be universal for all the ERE parameters.
For example, through concrete calculations, one could find that in
$^3S_1$ channel, $r_e$ is totally independent of $J_0$, $a$ is most
strongly influenced by $J_0$, while the rest of ERE parameters are
only weakly affected due to the suppression just mentioned. More
detailed analysis will be given in a forthcoming
report\cite{3s1-3d1}.

\begin{table}[t]
\caption{Naturalness/unnaturalness of ERE parameters in uncoupled
channels}
\begin{center}
\begin{tabular}{c|c|c}
\hline\hline
 Channels & natural  & (might be) unnatural   \\\hline
$^1S_0$  & $\{r_e, v_k, k\geq2\}$& $a$\\\hline $^1P_1, ^3P_0, ^3P_1
$& $\{a, v_k, k\geq2 \}$&$r_e$ \\\hline $^1D_2, ^3D_2$ &$\{a, r_e,
v_k, k\geq3\} $& $v_2$ \\\hline
 $\cdots(L\geq3)$ &$\{a, r_e, v_k, k\geq2, k\neq L\}$ & $v_L$ \\
\hline\hline
\end{tabular}
\end{center}
\end{table}

In earlier EFT treatments, the distinctive aspects of
nonperturbative renormalization demonstrated above were not fully
appreciated, leading to quite some debates\cite{BvKERev} (for recent
debates, see\cite{NTvK,PVRAB,EpelMeis}). A number of different
schemes were proposed in order to remove the 'inconsistency' of
Weinberg's power counting\cite{NTvK,KSW,BBSvK}, with some
"perturbative-like" expansion schemes being
advanced\cite{KSW,BBSvK}. As is pointed out above, the inconsistency
is in fact a misinterpretation of the failure of conventional wisdom
of renormalization. Specifically, nonperturbative counterterms do
not need to follow EFT power counting. Therefore, it is both
difficult and unnecessary to maneuver a unified power
counting\cite{NTvK,EpelMeis}. The entanglement property means that
the problems could well be resolved with appropriate choice of
nonperturbative prescriptions constrained by physical boundaries or
conditions. After all, the ultimate goal of any sensible scheme or
prescription should be to approach the physical dependence of
$T$-matrices upon $p$ as far as possible. Thus, a (new) formally
consistent power counting is not the full story: The nonperturbative
prescription must be appropriately defined to match physical
boundaries. For example, for a "perturbative-like" expansion scheme
to work, the following two criteria must be satisfied: (1) The
expansion converges; (2) Physical boundaries are fulfilled. Both
criteria are dependent upon prescription choice. To illustrate this,
we expand $1/T_{L}$ in Eq. (\ref{InvT-ll}) as follows: $$
\frac{1}{T_L}={\mathcal{I}_0}+ \frac{1+\delta N_L}{D_{L}^0 +\delta
D_L}p^{-2L}\simeq\frac{1}{T_L^0}+{\mathcal{O}} \left(\frac{\delta
N_L}{D_{L}^0}, \frac{\delta D_L}{D_{L}^0}\right)p^{-2L},$$ with
$T_{L}^0\equiv({\mathcal{I}_0}+ (1/D_{L}^0+\Delta_L )p^{-2L})^{-1}$
being the starting nonperturbative amplitude. Then, convergence
requires that $|T_{L}^0{\mathcal{O}} (\delta N_L/D_{L}^0, \delta
D_L/D_{L}^0)p^{-2L}|\ll1$, which in turn demands a sophisticated
renormalization prescription after the couplings are given. Next,
prescription must also be so chosen that $T_{L}$ fulfills physical
boundaries. Therefore, it is a challenging task to find a
prescription to fulfill the above two criteria.

Evidently, the informative form of the $T$-matrices will inspire new
investigations in the future. It would be interesting to explore the
relations between the nonperturbative parametrization elaborated
here and those in literature, for example, the subtractive
approaches\cite{YPhillips,EPVRAM,Frederico,ESoto}, and the lattice
approaches\cite{lattice}.

Now we conclude with the following remarks. In general, the ultimate
goal of a field theoretical calculation in nonperturbative regime is
to identify and parametrize all the elements that govern the
physical behaviors of the corresponding objects, especially the
elements hidden in divergences. To this end, we have achieved the
following: First, a fundamental parameter masked by a divergent
integral was identified and shown to be RG invariant and inherent in
all channels; Second, universal forms of nonperturbative
$T$-matrices with respect to prescription dependence were obtained
in all channels in the case of contact potentials; Third, within the
realm of contact potentials, a simple scenario led us to predict
that all the scattering lengths except those in the $S$-channels'
are natural, while higher ERE parameters like $r_e, v_k,k\geq2$
might also be unnatural in appropriate higher channels; Fourth, some
distinctive notions about nonperturbative renormalization were
revealed along with the failures of the conventional wisdoms,
providing a different resolution of the intriguing problem with
Weinberg's power counting in the EFT approach of $NN$ scattering.
These conceptual gainings are significant from purely theoretical
standpoint as nonperturbative renormalization is a challenging
issue. Finally, we stress again that the notions and conclusions
presented here are fairly general and hence illuminating for
nonperturbative treatments of any systems dominated by
short-distance interactions.

\section*{Acknowledgement}The author is deeply grateful to the
anonymous referees for their valuable comments that significantly
improved the presentation of our manuscript. The project is
supported in part by the National Natural Science Foundation of
China under Grant Nos. 10205004 and 10475028 and the Ministry of
Education of China.

\end{document}